\documentclass[11pt,preprint,flushrt]{aastex631}

\usepackage{natbib,graphicx,amsmath,subfigure,color}
\usepackage{verbatim}
\usepackage{soul} 

\usepackage{siunitx}
\def\eqq#1{Equation~(\ref{#1})}

\newcommand\ie{{\it i.e.\/}}

\newcommand{\vecx}{\mbox{\boldmath $x$}}

\newcommand{\vecu}{\mbox{\boldmath $u$}}

\begin{document}


\keywords{}
\title{Removing atmospheric turbulence from ground-based astrometry with fast correlation function measurements}

\author[0000-0001-6299-2445]{Daniel C. H. Gomes}
\email{dchgomes@sas.upenn.edu}
\affil{Department of Physics \& Astronomy, University of Pennsylvania, 
209 S.\ 33rd St., Philadelphia, PA 19104}

\author[0000-0002-8613-8259]{Gary M. Bernstein}
\affil{Department of Physics \& Astronomy, University of Pennsylvania, 
209 S.\ 33rd St., Philadelphia, PA 19104}

\author[0000-0002-7397-2690]{Claire-Alice H\'ebert}
\affil{Brookhaven National Laboratory, Physics Department, Upton, NY 11973, USA}

  \begin{abstract}
    We present a code that removes $\sim 90\%$ of the variance in astrometric measurements caused by atmospheric turbulence, by using Gaussian process regression (GPR) to interpolate the turbulence field from the positions of stars measured by Gaia to the positions of arbitrary targets. This enables robust and routine accuracy of 1--3~milliarcsec on bright sources in single exposures of the Dark Energy Survey (DES) and the upcoming Legacy Survey of Space and Time (LSST).
    For the kernel of the GPR, we use the anisotropic correlation function of the turbulent displacement field, as measured directly from the Gaia reference stars, which should yield optimal accuracy if the displacement field is Gaussian.
    We test the code on 26 simulated LSST exposures and 604 real DES exposures in varying conditions. The average correlation function of the astrometric errors for separations $<1\arcmin$ is used to estimate the variance of turbulence distortions. 
    On average, for DES, the post-GPR variance is $\sim 12 \times$ smaller than the pre-GPR variance.  Application of the GPR to LSST is hence equivalent, for brighter stars and asteroids, to having 12 Rubin observatories running simultaneously. The expected improvement in the RMS of turbulence displacement errors is the 
    square root of this value, $\sim 3.5.$
The post-GPR RMS displacement decreases with the density of reference stars as $\sim n_\star^{-0.5}$ for noiseless LSST simulations, and $\sim n_\star^{-0.3}$ for DES data.
\end{abstract}

\section{Introduction}

Humanity has long held a particular interest in astrometry—that is, keeping track of the positions and motions of celestial bodies. Archaeological sites with astronomical alignments show that several ancient civilizations marked the paths of the Sun, the Moon and other bodies, allowing us to consider astrometry as the oldest branch of astronomical science \citep{Dick2020}. Besides its historical roots, astrometry has also been fundamental to many practical endeavours, from the creation of calendars in antiquity to nautical and space navigation, and improvements in precision have consistently sparked astronomical discoveries \citep{Kovalevsky_2004}.

In the Hellenistic period, stellar positions were determined to an $\mathcal{O}$($\ang{1}$) precision from naked eye observations, an achievement traditionally attributed to Hipparchus (2nd century BC), but preserved only in the work of Ptolemy a few centuries later \citep{Dick2020}. During the Middle Ages and Renaissance, improvements were made possible with the construction of large measuring instruments that physically divided a circle into small angles, such as quadrants or sextants, leading up to Tycho Brahe’s observational precision of $\sim 20''$ \citep{brahe1598,Perryman2012}. With the invention of the telescope (17th century) and an increased interest in the use of astrometry for navigation, further gains in precision prompted the discovery of stellar proper motions \citep{halley}, stellar aberration \citep{Bradley} and the first parallax measurements \citep{Struve1837,Bessel1838,Henderson1840}.

From the end of the 19th century forward, photography allowed a faster and more permanent way of doing astrometry, and surveys eventually reached sub-arcsecond precision. Contemporary wide-area astrometric catalogs based in whole or part on ground-based electronic detectors include the United States Naval Observatory's UCAC4 \citep{UCAC4} Bright Star Catalog \citep{BSC}, and Pan-STARRS 1 catalog \citep{PS1}.
At this point, the Earth's atmosphere becomes the limiting factor for resolution and positional accuracy. Further great advances in accuracy were thus made by the astrometry-oriented space missions \textit{Hipparcos} \citep{Hipparcos}, then \textit{Gaia} \citep{GaiaDR3}.

Techniques to counter atmospheric effects in wide-area ground-based astrometry continued to be developed. \citet{2017Bernstein} demonstrated, for the Dark Energy Survey Camera, that the chromatic effects of the atmosphere and optics, detector distortions, and other deterministic elements of an astrometric calibration, can be reduced
to $\sim$ 1\,mas RMS level by calibrating using interlaced sky exposures. The remaining errors are correlated on small scales, with a coherence length of 5-10'. They further showed how this remaining displacement field must be originated by stochastic distortions induced by atmospheric turbulence, and that it has a typical RMS of $\sim 10$\,mas in the 90-second DES exposures.   \citet{Fortino_2021} tested the possibility of reducing such distortions using \textit{Gaia} stars to interpolate the turbulence field with Gaussian process regression (GPR). They obtained an order-of-magnitude reduction of the variance of the displacements of hundreds of Dark Energy Survey exposures. \citet{Leget} demonstrated similar gains from GPR interpolation of astrometry in exposures from the HyperSuprimeCam (HSC) survey.
We explore the same GPR idea, resolving certain scalability and stability issues that were present in the previous works, and yielding a code that is practical for use in upcoming surveys such as the Legacy Survey of Space and Time (LSST) as well as retrospective recalibration of past surveys.

\section{Methods}

\subsection{Gaussian process interpolation}
We consider that each star at a location $\vecx$ has an astrometric displacement $\vecu$ due to atmospheric turbulence, and this displacement can be described as a gaussian random field, such that
\begin{equation}
\vecu \sim \mathcal{N}(\boldsymbol{\mu},K)
\end{equation}
where $\boldsymbol{\mu}$ is the mean and $K$ is a covariance matrix constructed from a kernel function $K(\vecx_i,\vecx_j) = K(\vecx_i-\vecx_j)$. Since we remove constant and low-order polynomial terms from our displacement fields, we can assume $\mu = 0$.
A Gaussian process regression (GPR) will use the positions $\vecx$ and displacements $\vecu$ of a set of N reference stars  to infer the displacements $\vecu '$ on a set of M targets at $\vecx'$. Let us define:
\begin{itemize}
\item $\mathbf{K}$ as the $N \times N$ covariance matrix between reference stars,
\item $\mathbf{K'}$ as the $N \times M$ matrix of covariances between reference and target stars,
\item $\mathbf{K''}$ as the $M \times M$ covariance matrix between target stars.
\end{itemize}
The GPR solution is derived \citep{Rasmussen2005} from the joint Gaussian distribution:
\begin{equation}
\begin{pmatrix}
\vecu \\
\vecu '
\end{pmatrix}
\sim \mathcal{N}\left(
0,
\begin{bmatrix}
\mathbf{K} & \mathbf{K'} \\
\mathbf{K'}^{T} & \mathbf{K''}
\end{bmatrix}
\right).
\end{equation}
\begin{equation}
\label{gp_estimator}
\Rightarrow \ \vecu ' \mid \vecu \sim \mathcal{N} \left(
\mathbf{K'}^{T} \mathbf{K}^{-1} \vecu,\;
\mathbf{K''} - \mathbf{K'}^{T} \mathbf{K}^{-1} \mathbf{K'}
\right).
\end{equation}
The estimated astrometric displacements for the target stars correspond to the mean of the distribution: $\mathbf{K'}^{T} \mathbf{K}^{-1} \vecu$.
A solution can be found via singular value decomposition,.
\begin{equation}
\label{svd}
\mathbf{K} = \mathbf{U} \, \mathbf{S} \, \mathbf{V}^{T}.
\end{equation}
Since, in this case, $\mathbf{K}$ is a symmetric, positive semi-definite matrix, the matrices $ \mathbf{V}$ and $\mathbf{U}$ can be chosen to be equal and the singular values of $\mathbf{K}$ are the eigenvalues, given by the diagonal entries S of the matrix $\mathbf{S}$. Then,

\begin{align}
\label{prepare_gp}
\mathbf{z} &= \mathbf{U}^{T} \vecu \\
\label{prepare_gp_2}
\mathbf{w} &= \mathbf{S}^{-1} \circ \mathbf{z} \\
\label{prepare_gp_3}
\boldsymbol{\alpha} &= \mathbf{V} \mathbf{w}
\end{align}
where $\circ$ in \eqref{prepare_gp_2} represents element-wise multiplication. The mean at the target point locations is finally obtained as
\begin{equation}
\label{apply_gp}
\mathbf{u'} = \mathbf{K'}^{T} \boldsymbol{\alpha}.
\end{equation}

The diagonal elements of the second term in parentheses in \eqref{gp_estimator} will provide the variance of our estimates at each target point,
\begin{align}
\label{variance_gp}
\mathbf{T} &= \mathbf{U}^{T} \mathbf{K'}, \\
\label{variance_gp_2}
\operatorname{Var}[u'_i] &= K_0 - \sum_{j=1}^{N} T_{ji}^2 \cdot (S^{-1})_j,
\end{align}
where $K_0 = K(\vec{0})$ is the zero-lag variance -- i.e., the value the kernel assumes for $\vecx_i = \vecx_j$.
The computation time should be dominated by the $O(N^3)$ operations it takes to execute the SVD in \eqref{svd}; however, if $M>N$ then the application to the target points in \eqref{variance_gp}, which takes $O(N^2M)$ operations, can become the bottleneck.  If we were to demand the full covariance matrix of the target values rather than only the diagonal elements, this would slow down the process considerably and require problematically large storage for the resultant catalog.
\subsection{Fast correlation function measurements}
The GP kernel, as a descriptor of the similarity between the turbulence displacement field in two different points, determines which combination of reference stars is used for field estimation in each target location, and how they should be weighted. A simple solution that bypasses the need for an actual interpolation is to consider the nearest $N$ reference stars and average the field at their locations, as done by \citet{Lubow_2021} for the Pan-STARRS1 Data Release 2. Moving towards more accurate correlation tracking, \citet{Fortino_2021} assume a functional form for the kernel, and use an optimization routine for the kernel parameters, starting from an estimate based on their training set, and varying them until minimal RMS errors are achieved on the validation set. This step, however, makes their routine computationally expensive. On average, a GPR is re-calculated $\sim 100$ times, which creates scalability challenges in larger catalogs such as the expected LSST data. 

In principle, the actual correlation function of the $\vecu$ field is the optimal GPR kernel. The fast correlation measurement code \texttt{TreeCorr} \citep{Jarvis2015} produces an empirical correlation function of the difference between DES and Gaia positions of the reference stars.  
\citet{Leget} noted that using the (noisy) measured correlation function directly in the GPR can lead to a correlation matrix $\mathbf{K}$ that is not positive-definite, invalidating the GPR.  Their solution is to fit a parametric functional form to the measured correlation function, which is then used as the kernel.  We pursue a different method, described below, that uses the empirical correlation function but eliminates the non-positive eigenvalues of $\mathbf{K}$.
 
 The correlation-function measurement can use weights that favor reference stars with smaller nominal measurement errors, since these will be more reliable estimates of the turbulence field at that location. A Gaussian process kernel derived from this measurement does not have to undergo parameter optimization, and therefore is considerably faster than the method of \citet{Fortino_2021}. Section \ref{compute} will show computational time estimates for our full pipeline and the dependence on the number of reference stars.

\subsection{Joint $u$ and $v$ inference}
\label{jointuv}
When performing the GPR, we can choose between processing the $u$ and $v$ fields separately, or joining them in a single correlation matrix.
\begin{equation}
K_{\text{joint}} = \begin{bmatrix}
K_{uu} & K_{uv} \\
K_{uv} & K_{vv}
\end{bmatrix}
\end{equation}

The joint $uv$ solution requires inverting a single matrix of size $2N$ instead of 2 matrices of size $N,$ which is nominally $4\times$ more operations. The potential advantage of this slower method is that it does not ignore the cross-correlation between the $u$ and $v$ fields. Therefore, it is only worthwhile if that cross-correlation provides sufficiently informative constraints on the turbulence patterns.

\subsection{Data for testing}

We test our method on two different sets of data. The first consists of a set of 26 simulated displacement fields from atmospheric turbulence patterns expected to mimic those encountered in the LSST data taken at the Rubin Observatory.
The simulations follow the model described in \citet{Hebert2024}, where displacements are assumed to originate from discrete layers of turbulent air (one for the ground layer and five for the free atmosphere), each modeled as a screen of von K\'arm\'an turbulence \citep{vonkarman1948}. The parameters setting the turbulence power spectrum and dynamics of the turbulence screens are inferred from observations near the Rubin site. 

The second set are 604 DES exposures, chosen from four different sky zones, where the matching with actual Gaia DR3 sources was performed, and turbulence displacements were derived via GPR for the remaining (target) stars.

\section{Code}
\label{code}
For a single DES exposure, our code performs the following steps to generate updated turbulence-subtracted position estimates and their uncertainties. All input values explicitly mentioned below are default values, which can be changed by passing a different value as an argument to the corresponding function.
\begin{enumerate}
\item \textbf{Load data:} The \textsc{SExtractor} \citep{sextractor} object catalog produced from this single exposure catalog is loaded as well as the catalog produced from a coaddition of many exposures covering the region. A cut on the \textsc{SExtractor} paramater \texttt{SPREAD\_MODEL} in the coadd catalog
 removes detections inconsistent with a point source. We load a table with information on each DES exposure such as MJD, coordinates of the field center, and ICRS observatory position. A gnomonic projection to coordinates $(x,y)$ about the field center is applied to the RA and Dec coordinates from both catalogs. A k-d tree is used to find all coadd objects that are uniquely within 1 arcsecond of each single-exposure source. These are considered matches, and the $g-i$ color of the coadd object is assigned to the observed source. A default color can be assigned for sources with no matches, in case we want to predict the turbulence at their locations as well. The color is then used as an input to refine the single-exposure sky coordinate values via the astrometric calibration solutions for DECam derived by \citet{gbdes}. The updated coordinates are re-projected around the center of the exposure, and the new $(x,y)$ positions are stored in a table. We also keep track of which objects were successfully matched with the coadd colors and which ones had a default color assigned to them.

\item \textbf{Gaia catalog matching:} We load the Gaia DR3 solutions for all sources within the region covered by the exposure. Sources with no parallax or proper motion information are ignored. We use the full solutions to model the position of each source at the time of the DES exposure, and perform a gnomonic projection of these modeled positions around the DES pointing coordinates. Then, we build a k-d tree and find which Gaia stars are uniquely within a certain match radius of the DES source. The stars with successful matches are considered reference stars, as long as they satisfy the two following conditions: the nominal astrometric error from the DES single-exposure catalog is below a threshold pre-defined by the user, and the source was previously matched to a coadd source (and therefore has a reliable color estimate). An initial estimate of the turbulence at the position of each reference star is computed by subtracting the position predicted by Gaia from the DES position. We denote the field of displacements in the E-W direction as $u$; and in the N-S direction as $v$. All remaining stars are considered targets, i.e., points for which we want a turbulence correction prediction.

\item  \textbf{Polynomial fit:} The largest scales of correlation in the telescope field of view may be associated with systematic errors, and are best captured by low order polynomial fits. Removing these scales allows the GPR to better capture the smaller scales, which are characteristic of turbulent distortions. We perform a third-order polynomial fit of the displacements at the position of the reference stars. The input uncertainty for the fit is a combination of the nominal astrometric error in each position and a fixed floor set at the expected RMS of turbulent distortions. The resultant third-order polynomial is stored as an initial component of the turbulence model. In order to remove outliers from the reference set, sources on the reference set that deviate more than $3\sigma$ from the polynomial fit are removed from the reference set and reassigned as targets. This clipping of outliers can be applied several times, with new fits done without the outliers, until there are zero sources above $3\sigma$, or until a pre-assigned maximum number of rounds.

\item \textbf{Build kernels:} 
\label{Cstep}
The 2d correlation function of the polynomial-subtracted displacements of the reference stars is measured using TreeCorr \citep{Jarvis2015}. The maximum separation in each coordinate is set to 0\fdg5, in bins of width 20\,\arcsec, rendering each correlation function as a 180x180 pixel grid that has inversion symmetry about the origin but is not circularly symmetric. We form the correlation functions $C_{uu}$ and $C_{vv}$ of the two displacement coordinates, plus the $C_{uv}$ cross-correlation if using the GPR variant described in Section~\ref{jointuv}.
We then process these measured correlations as follows to suppress measurement noise. First, we bin blocks of $2\times2$ pixels in such a way that $C(0,0)$ is located at the center of the central pixel. Second, we multiply each correlation function by a Blackman-Harris apodization function, so that the correlation goes smoothly to zero at the predefined maximum separation, which suppresses noise in the Fourier transform. Lastly, we take the Fourier transform and filter the power spectra: For the $uu$ and $vv$ correlations, we keep only the Fourier components that are above a certain signal-to-noise threshold (default at 2.5), where the noise is estimated from scatter of power at the higher half of frequencies. Elements with signal-to-noise below the threshold are set to zero. We adjust the $uv$ power spectrum to ensure the $2\times2$ covariance matrix at each $k$ is non-negative definite, by clipping $P_{uv}(k)$ to be between 0 and $\text{sign}(P_{uv}(k))\sqrt{P_{uu}(k)P_{vv}(k)}.$ Finally, the power spectra are Fourier transformed back to correlation functions and stored. 
\item \textbf{Divide sources in patches:} Since any turbulent or systematic astrometric-error correlations on angular scales $\gg 0\fdg1$ are weak and easily removed from target stars with knowledge of the $O(100)$ reference stars within 0\fdg1, there is no gain in accuracy from performing a single GPR over the entire field of view (FOV) of a DES or LSST exposure. It would in fact be counterproductive, because the required matrix inversion has a complexity of $\mathcal{O}(N^3)$ for $N$ reference stars used, and the matrix multiplication to obtain the solution for $M$ targets adds $\mathcal{O}(N^2M)$. It is faster by a factor of $P^2$, therefore, to perform the GPR operations on $P$ smaller patches each with $1/P$ fewer reference and target stars, as long as the patches are large enough to capture the largest scale of relevant correlations. We divide the DECam FOV into 10 patches, each with sides of length $\sim35-40\arcmin$, and perform the following step for each of them. Only target stars strictly within the patch are selected, but we include reference stars within a certain buffer distance (defaulting to 6\arcmin), so that targets near the border of the patch have enough relevant reference stars for an accurate solution.

\item \textbf{GPR and cross-validation:}
Each GPR run follows equations \eqref{svd}-\eqref{variance_gp}. The correlation matrices $K,K^\prime, K^{\prime\prime}$ for each star pair are interpolated from the correlation function computed in Step~\ref{Cstep}.
 To each diagonal element $K_{ii}$ or $K^{\prime\prime}_{ii}$ we add the photon-noise variance in the position reported by the measurement software, and the variance reported by Gaia, if relevant.
 
The SVD defined in \eqq{svd} allows us to clip the singular values to a chosen (positive) minimum value. This prevents numerical instabilities that could result from noise in the measured $C_{uu}$ and $C_{vv},$ and prevents stars with abnormally or anomalously low stated measurement errors from dominating the solution.
This mechanism was not necessary for the exposures we utilized, and if this remains true in future applications, it might be possible to accelerate the GPR by using a Cholesky decomposition instead of an SVD.
 
 We can reserve a subset of the reference stars from the GPR as a validation set. These reserved stars are treated as targets in the GPR algorithm, so their post-GPR model-subtracted residuals from Gaia data can be computed and used to quantify the performance of the algorithm. 
Futhermore, if we perform a full $N$-fold cross validation---dividing the reference stars into $N$ disjoint sets, and performing $N$ GPRs each reserving one of these groups as mock targets and the remaining ones as reference stars---then every reference star is reserved once and has a turbulence estimate that is independent of its own data.
The number of cross-validation runs can be defined by the user (the default is $N=5$). The actual target stars can be ignored during cross-validation.
Reference stars whose cross-validated positions are 
$>3\sigma$ deviant from their Gaia data are excluded as outliers for the final GPR run on the actual targets.

\item \textbf{GPR with targets:} The entire set of reference stars is used to perform a final GPR and compute the predicted turbulence at the target locations. 
 To avoid memory issues on exposures containing dense star fields, we can perform the GPR for a patch in two steps: an initial step applies equations \eqref{svd}--\eqref{prepare_gp_3}, which depend exclusively on the reference stars.  In the second step, we divide the targets into batches of a specified maximum size, and for each batch compute the final GPR prediction following equations
\eqref{apply_gp}--\eqref{variance_gp_2}.

\item \textbf{Compute statistics:} We compute statistics of the reserved stars' pre- and post-GPR residuals from Gaia to estimate the improvement on turbulence-related astrometric errors. The simple RMS of the reserved-star residuals is inadequate for this, since it is a mixture of turbulence errors and image-measurement errors.  Since image noise is uncorrelated between stars, we can isolate the turbulence effects by measuring the correlated errors between close pairs of stars.
We then compute a total for the $u$ and $v$ fields, \ie
\begin{equation}
\label{prepost}
C_0 \simeq \langle C_{uu}(r<1')\rangle+\langle C_{vv}(r<1')\rangle
\end{equation}
The value of $r=1\arcmin$ for the variance statistic $C_0$
is a compromise between two problems: 
the correlation functions $C(r)$ are decreasing with $r,$ so the larger $r$ is chosen, the more  \eqq{prepost} will underestimate the true variance $C(0,0).$  On the other hand, the smaller we make $r,$ the fewer pairs of reserved stars are available, and the noisier our estimate of $C_0$ becomes.  The ratio between the $C_0$ values of the original measurements vs the GPR-corrected positions is our metric of turbulence suppression.

\item \textbf{Save results:} A table is saved with the original observed coordinates and the model displacements, with their respective errors, for both reference and target stars. Model-subtracted residuals are also stored for the reference stars, as well as a column indicating whether the target had a color estimate from the coadd catalog or whether a default value was used. The summary statistics for each exposure are also stored. 

\end{enumerate}

When applying the code on LSST simulated data sets, where the true turbulence displacement is available for all sources, no matching with a real Gaia catalog is needed. In this case, we start at step 3. We perform the GPR in patches of similar size,  $\sim35-40\,\arcmin$, to those used for the DECam exposures.

\section{Results}
\subsection{LSST}
For the 26 simulations of LSST exposures with different intensities of turbulence displacements, we have the capability to measure how
the GPR improves with an increasing density of reference stars beyond the local Gaia magnitude/density limit. First we investigate this for ``noiseless'' stellar positions, i.e., when the additional fake stars are bright enough to have their measurement noise set at a $1$\,mas lower bound, and then for a second scenario, 
in which stellar density is increased through addition of fainter stars, and their realistic measurement noise is applied to the stellar positions.  In the second scenario, the added reference stars would be fainter than Gaia's limit, potentially arising from 
5-parameter solutions from LSST itself, and thus they would have higher per-exposure measurement errors. To simulate this scenario, we extract a distribution of LSST g-band magnitudes simulated by \citet{trilegal}, and draw magnitudes from this distribution to assign to the simulated stars. Astrometric errors are then predicted for each star from its magnitude and assuming average sky conditions computed from the \texttt{baseline\_v4.0\_10yrs.db} LSST simulation,\footnote{Available on \href{this site.}{https://survey-strategy.lsst.io/baseline/index.html}.} as detailed in \citet{Gomes_2023}. The error estimates are added in quadrature to the reference star correlation matrices during the GPR, as described in section \ref{code}.

\begin{figure}
  \centering
  \includegraphics[width=0.8\textwidth]{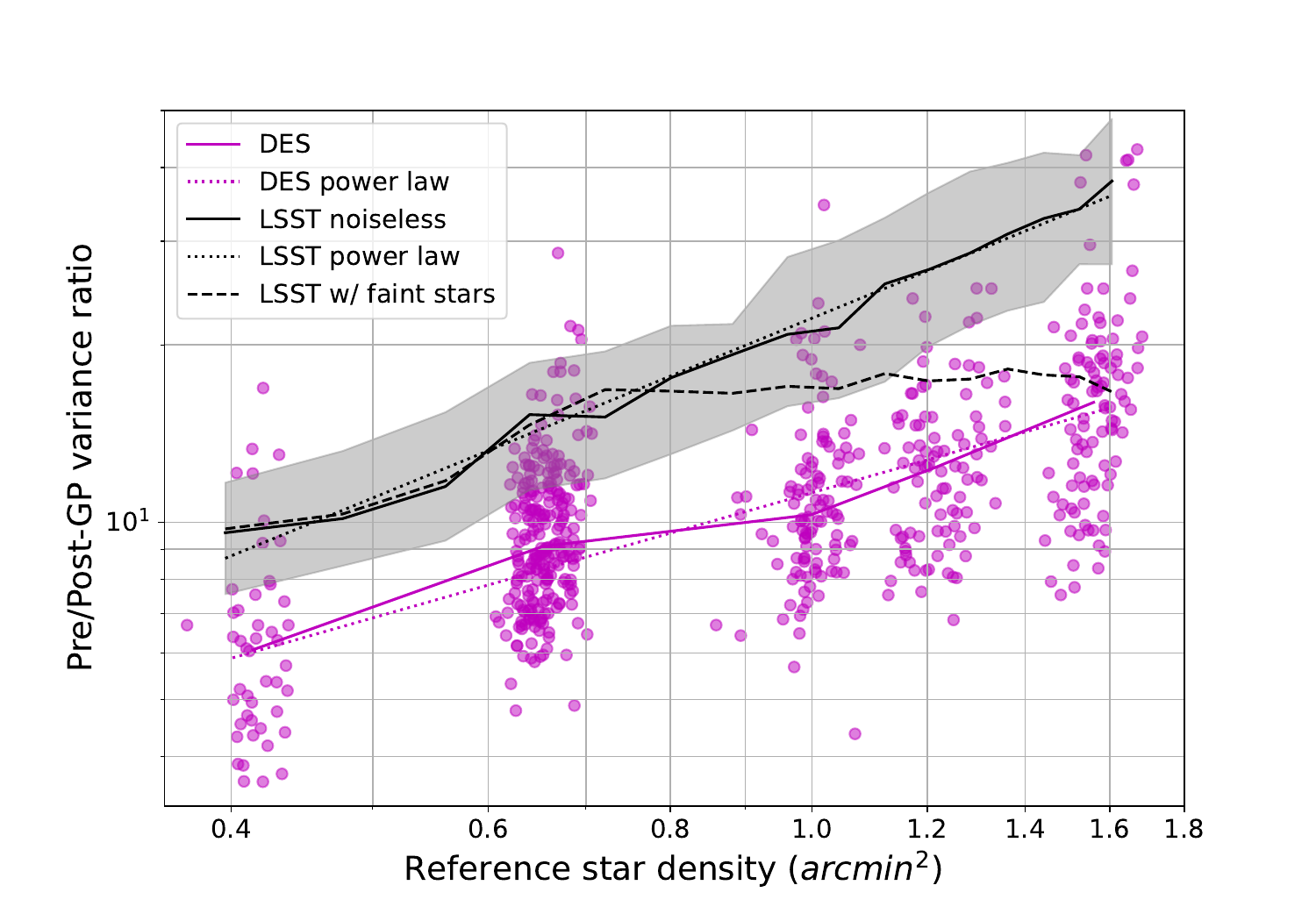}
  \caption{The turbulence variance reduction factor ($C_{0}^{ \text{pre}}/C_{0}^{\text{post}}$) is plotted as a function of the reference star density on LSST simulations (black) and real DES exposures (purple). The solid black shows the median variance reduction attained for LSST simulations with a fixed 1\,mas measurement uncertainty per star. The gray zone represents $1\sigma$ exposure-to-exposure RMS variation. The simulations shown with black dashed line also include expected magnitude-dependent LSST measurement noise per star, with higher density values representing the use of fainter-than-Gaia reference stars. The black dotted line is a power law fit to the variance reduction in the noiseless scenario. The magenta dots are individual DES exposures; the solid line, their median value, and the dotted line, a power law fit.}
  \label{fig:LSST_ratio}
\end{figure}

Figure \ref{fig:LSST_ratio} shows the median ratios between the model-subtracted post-GPR variance and the pre-GPR value for different densities of reference stars, in both the noiseless scenario and the one with realistic error estimates. Using the noiseless medians to perform a power-law fit, we find that the variance ratio scales with $\sim~n_\star^{1.0}$. The RMS residual turbulence per axis is estimated as the square root of our variance estimator after it is divided by 2 (to account for the two axes),
\begin{equation}
\label{RMS}
    \text{RMS} =  \sqrt{\frac{\langle C_{uu}(r<1')\rangle+\langle C_{vv}(r<1')\rangle}{2}}.
\end{equation}
For this noiseless scenario, it decreases with $\sim~n_\star^{-0.5}$. 

The realistic scenario reaches a plateau around $0.8\,\text{star}/\text{arcmin}^2$ (that is, an exposure with $1\,\text{star}/\text{arcmin}^2$ where the metrics are obtained via 5-fold cross-validation), beyond which point the photon noise errors of the new, fainter stars become too large to give information on the $1-3$\,mas post-GPR residuals left behind by the brighter stars. The magnitude limit associated with this threshold is around $21.5$ for the LSST g-band, and at this point the measurement errors are typically $\sim 15 \, \text{mas}$. This suggests LSST will gain rather little from using reference stars much fainter than Gaia's limit of $G \sim 21$.

In Figure \ref{fig:preposthist}, we restrict ourselves to the noiseless simulations with reference densities between 0.7 and 1.7\,$\text{star}/\text{arcmin}^2$ ($\times 0.8$ for effective grid density during cross-validation), and we plot the distribution of pre and post GPR variances $C_0$ to compare with our DES results. This selection of densities and use of the noiseless simulations is intended to reproduce the scenario where all reference stars come from Gaia. We exclude from this plot the simulations where at least one of the turbulence axes has variance larger than $500 \, \text{mas}^2$.

\subsection{DES}

\begin{figure}
  \centering
  \includegraphics[width=0.7\textwidth]{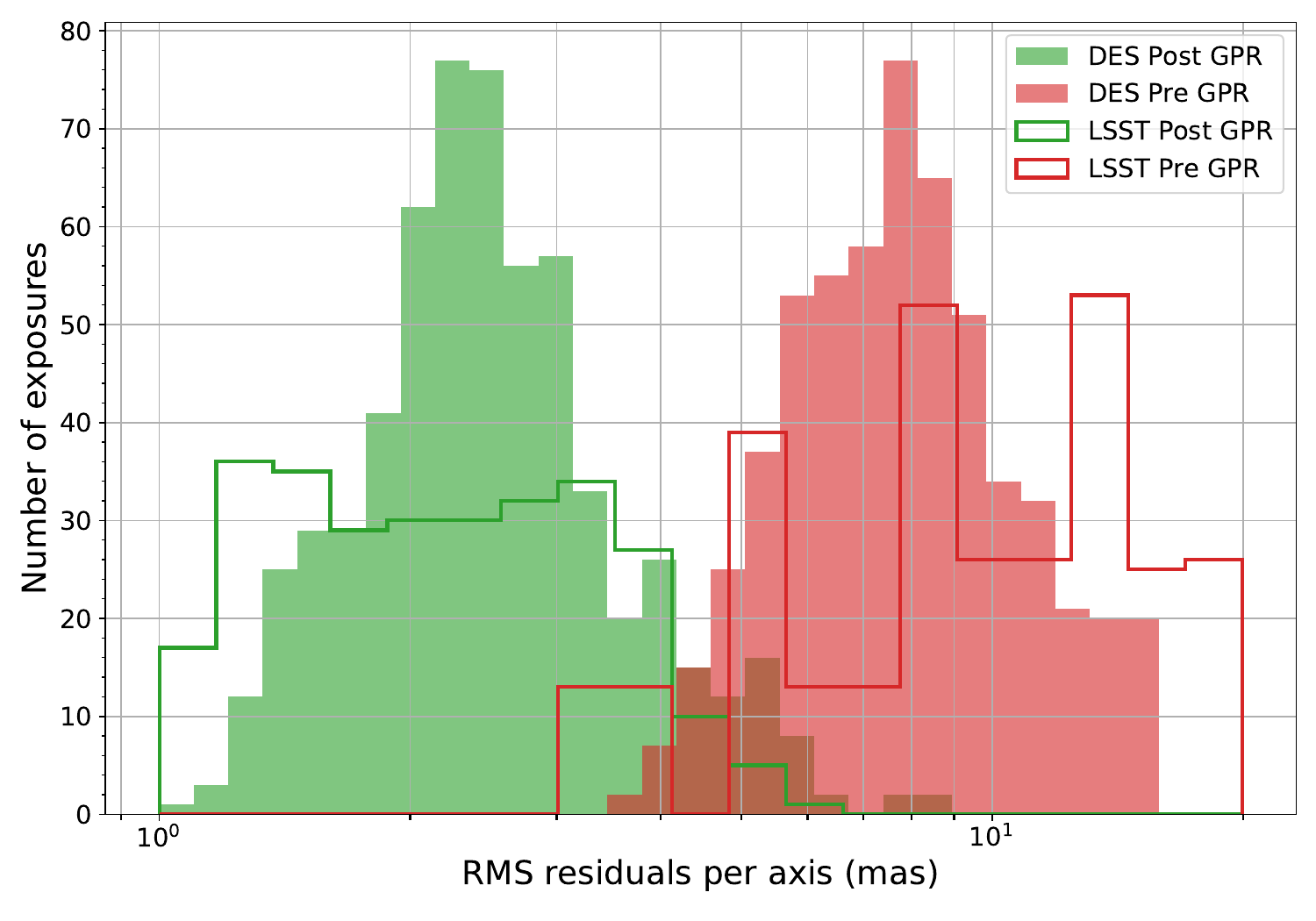}
  \caption{Distribution of RMS astrometric residuals per axis on all 604 DES exposures (filled) and LSST simulations (step) under similar conditions. Red denotes the RMS computed from the original correlation function of the displacements (pre-GPR); and green, the equivalent value obtained after subtraction of the GPR model (post-GPR).}
  \label{fig:preposthist}
\end{figure}

We apply our GPR code to 604 DES exposures. After subtraction of our GPR turbulence model from the measured positions of the reference stars, we consistently obtain an $\mathcal{O}(10)$ times reduction in the variance of the displacement field, as estimated by \eqq{prepost}.  We then compute the RMS per axis defined on \eqq{RMS}.
Recall that this estimator is likely to be biased somewhat low compared to the true RMS error.  Nonetheless Figure~\ref{fig:preposthist} shows that the turbulence RMS is reduced from 5--20\,mas to 1--4\,mas for the great majority of DES exposures. The LSST simulations show a similar range of values to the DES exposures, for both pre and post GPR variances. This is probably the result of two counteracting differences between LSST and DES: LSST will operate with a mirror twice as large, but the simulations assume 30 second exposures as opposed to the 90 second ones from the DECam. The mirror improvement would be equivalent to treating each turbulence patch as an average over 4 DECam patches; partially countered by the 1/3 decrease in exposure time.

\begin{figure}
  \centering
  \includegraphics[width=0.8\textwidth]{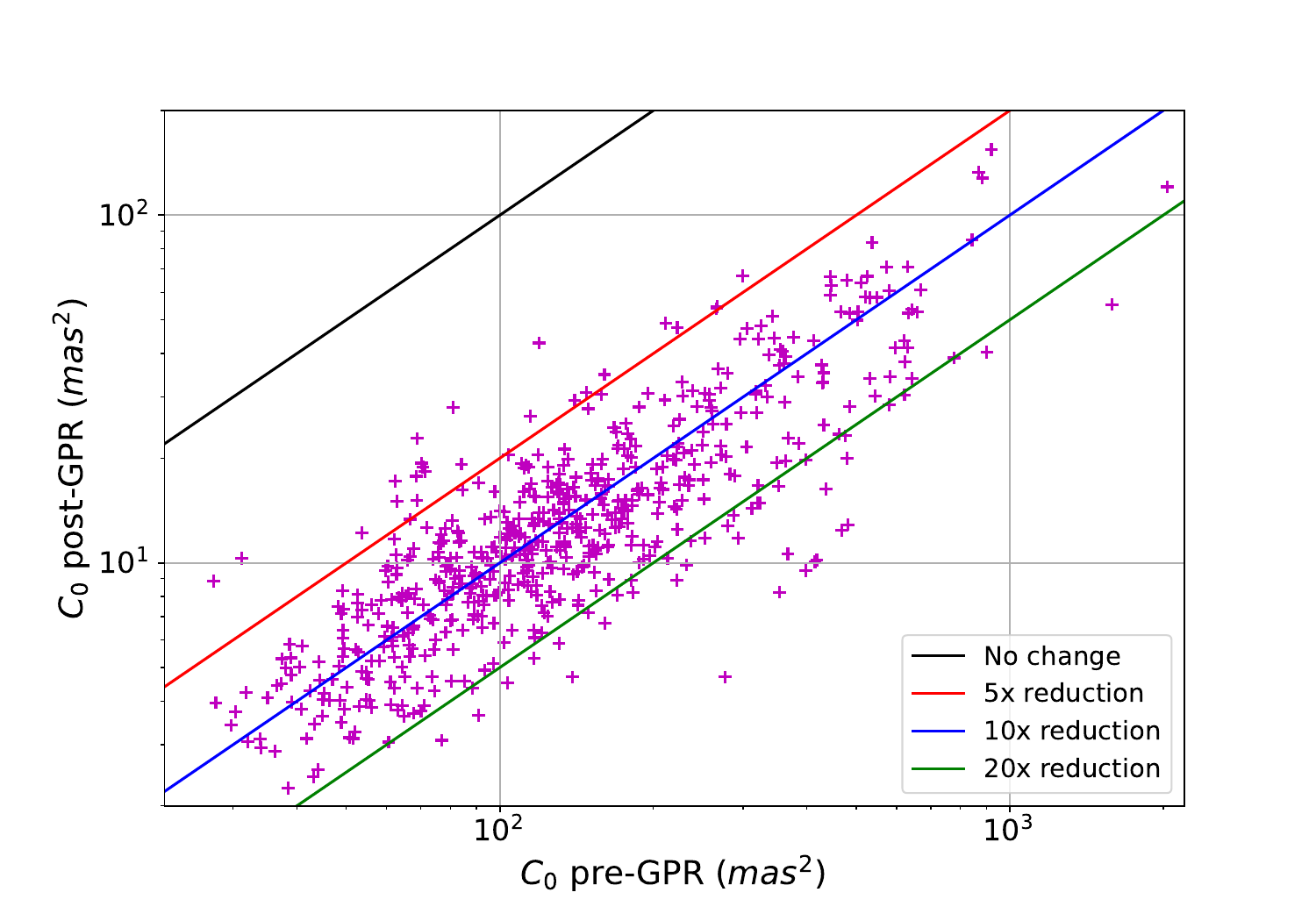}
  \caption{Improvement in the variance of the turbulence field on 604 DES exposures. The x-axis shows the pre-GPR variance computed from \eqref{prepost}; and the y-axis, the variance of the GPR-model-subtracted residuals. Each point corresponds to a single DES exposure. The diagonal lines depict the ratios: 1x (No change), 5x reduction, 10x reduction and 20x reduction, which correspond to 1x, 2.2x, 3.2x and 4.5x reduction of the per-axis RMS.}
  \label{fig:DES_res}
\end{figure}

Figure~\ref{fig:DES_res} plots pre- vs post-GPR variance $C_0$ for all of the DES exposures. The average pre/post ratio for all $grizY$ exposures is 12.2, which corresponds to a $\sim 3.5$x reduction of the RMS per axis. The dependence of the DES pre- vs post-GPR turbulence variance on the local Gaia star density is shown on Figure \ref{fig:LSST_ratio}, alongside the LSST predictions. A power-law fit shows that the variance ratio in DES increases $\sim n_{\star}^{0.7}$, which is smaller than the optimistic $\sim n_{\star}^{1.0}$ behavior seen in the noiseless LSST simulations. The post-GPR RMS per axis for DES decreases as $\sim n_{\star}^{-0.3}.$

The joint inference method described in Section~\ref{jointuv} does not yield any improvement compared to the separate inference of $u$ and $v$. The latter approach is therefore preferred due to its faster run time.

\subsection{Computational cost}
\label{compute}
\begin{figure}
  \centering
  \includegraphics[width=0.8\textwidth]{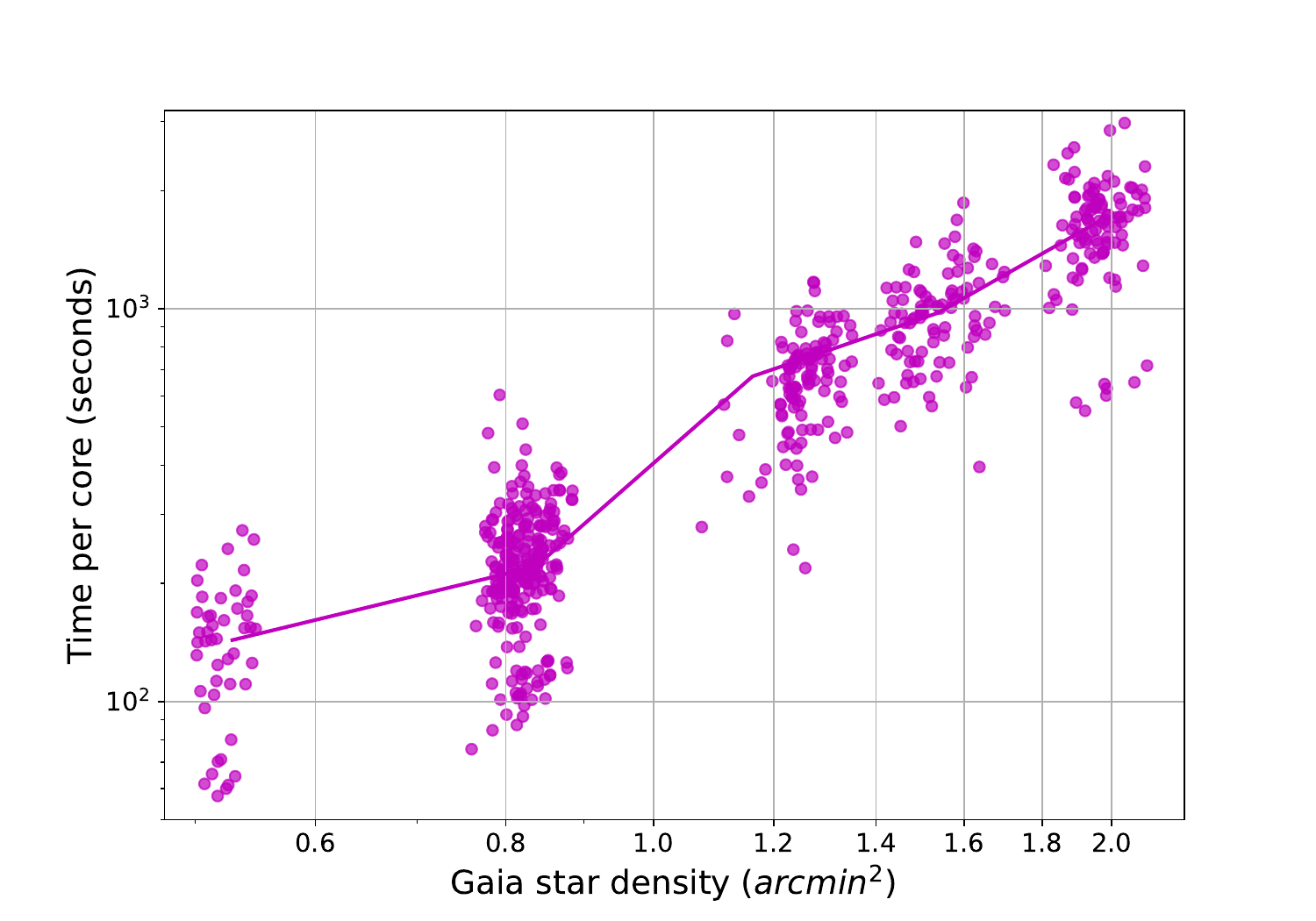}
  \caption{Time elapsed for a single CPU core to perform the routine GPR in an entire DES exposure divided in 10 patches, including, for each patch, five-fold cross-validation ($80\%$ of Gaia stars as reference) and a final run with all Gaia stars and the additional non-Gaia targets.}
  \label{fig:Compute}
\end{figure}

We track the time spent to perform the entire GPR routine on each DES exposure, if running with no parallelization on two CPU cores. The computational cost is shown in Figure \ref{fig:Compute}. We see that for a typical reference star density of $0.8 < \rho < 1.2\,\text{star}/\text{arcmin}^2,$ the exposure is processed in $\mathcal{O}(10^2)$\,core-seconds. This includes the five cross-validation regressions for each of the 10 patches in which the DECam focal plane is divided, using $80\%$ of the reference grid, and a final run for each patch where the solutions are applied to all targets. What dominates the runtime is not the $\mathcal{O}(N^3)$ matrix inversion, but rather the $\mathcal{O}(N^2M)$ multiplication, due to the fact that the number of targets $M$ greatly exceeds that of the reference stars. Accordingly, we can see the computational cost rising with $N^2$ in Figure \ref{fig:Compute}.

Based on these data, we estimate code performance for LSST processing, assuming that we are interested in exposures far from the galactic plane. For a Gaia star density of 1\,$\text{star}/\text{arcmin}^2$, a power law fit to our DES results shows a cost of $\sim 430$\,core-seconds for 10 patches. If we maintain the $35-40\,\arcmin$ patch size, the LSST focal plane can be covered with 25 patches, each with performance comparable to a single DES patch. The total exposure processing time would therefore be $\sim 2.5 \times 430$\,s $\simeq 18$\,core-minutes.

An average night with $\sim 500$ new griz LSST exposures (see \texttt{baseline\_v4.0\_10yrs.db}) would thus require 9000 minutes-cores for turbulence computation. This means a minimum of 6-7 cores would be needed to process one night of observations within 24 hours, ensuring no accumulation of unprocessed data. It is important to note that this assumes all exposures are in the optimal low-density regions. A 10-fold increase in source density applied to both reference and target sets translates into an 1000 $\times$ increase in computational cost, bringing up the minimum number of cores to the order of thousands. A possible way to limit this increase is setting a threshold to the number of reference stars in overdense regions, which is justified by the fact that all the reported variance improvements were attained with reference networks of $< 2$\,stars$/\text{arcmin}^2$.

\subsection{Further applications}

We look at the potential use of our code to improve both LSST minor planet astrometry and stellar proper motion estimates. First, we consider a set of $\sim 1000$ main belt asteroid orbits and match their orbits to the fields contained in the \texttt{baseline\_v4.0\_10yrs.db} LSST simulation. For each simulated observation, we compute the estimated astrometric errors in two scenarios: with and without turbulence reduction. The total error is a combination of the photon-noise error and a fixed astrometric floor that accounts for turbulence and remaining calibration errors. We set the astrometric floor at the pre-GPR nominal value of 10\,mas per axis for the uncorrected scenario, and at the post-GPR value of 3\,mas for the corrected scenario---a rough estimate considering the 10-fold expected variance reduction and the performance of the simulated LSST fields with similar level of pre-GPR variance.

We then compare both outcomes using Equation \eqref{information} to measure the positional information on each orbit from the entire data set.
\begin{equation}
\label{information}
I = \sum_i \frac{1}{\sigma_i^2}
\end{equation}

\begin{figure}
  \centering
  \includegraphics[width=0.8\textwidth]{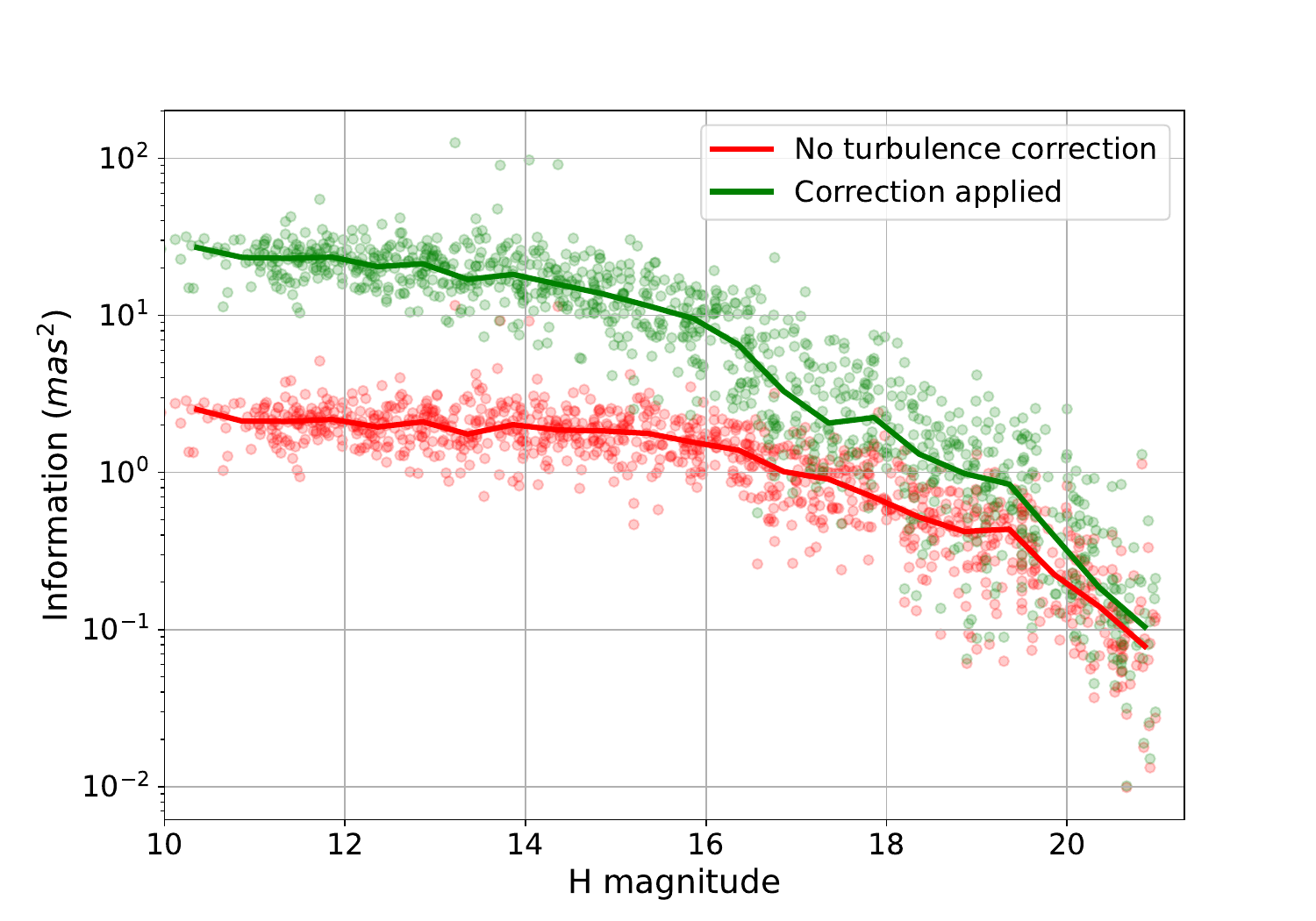}
  \caption{Total positional information expected after 10 years of LSST, for a sample of ~1000 Main Belt Asteroids. Red points denote the scenario where no turbulence correction is applied (10\,mas errors added in quadrature to the photon noise error). Green points reduce the astrometric errors according to the performance of our turbulence reduction code (the astrometric floor is moved from 10\,mas to 3\,mas). Continuous lines are medians.}
  \label{fig:mba}
\end{figure}

Figure \ref{fig:mba} compares the positional information for each object after the 10-year survey in both scenarios (corrected and uncorrected). Since fainter asteroids have a larger contribution of photon-noise error, they are expected to benefit less from turbulence reduction. We find that the total positional information on MBA orbits improves by a factor of $\approx 10 \times$ for the brightest bodies, and that this factor starts to reduce significantly between H = 16 and H = 18. For $H > 20$, the corrected and uncorrected samples have roughly the same information value, showing almost no benefit from turbulence reduction.

\begin{figure}
  \centering
  \includegraphics[width=0.8\textwidth]{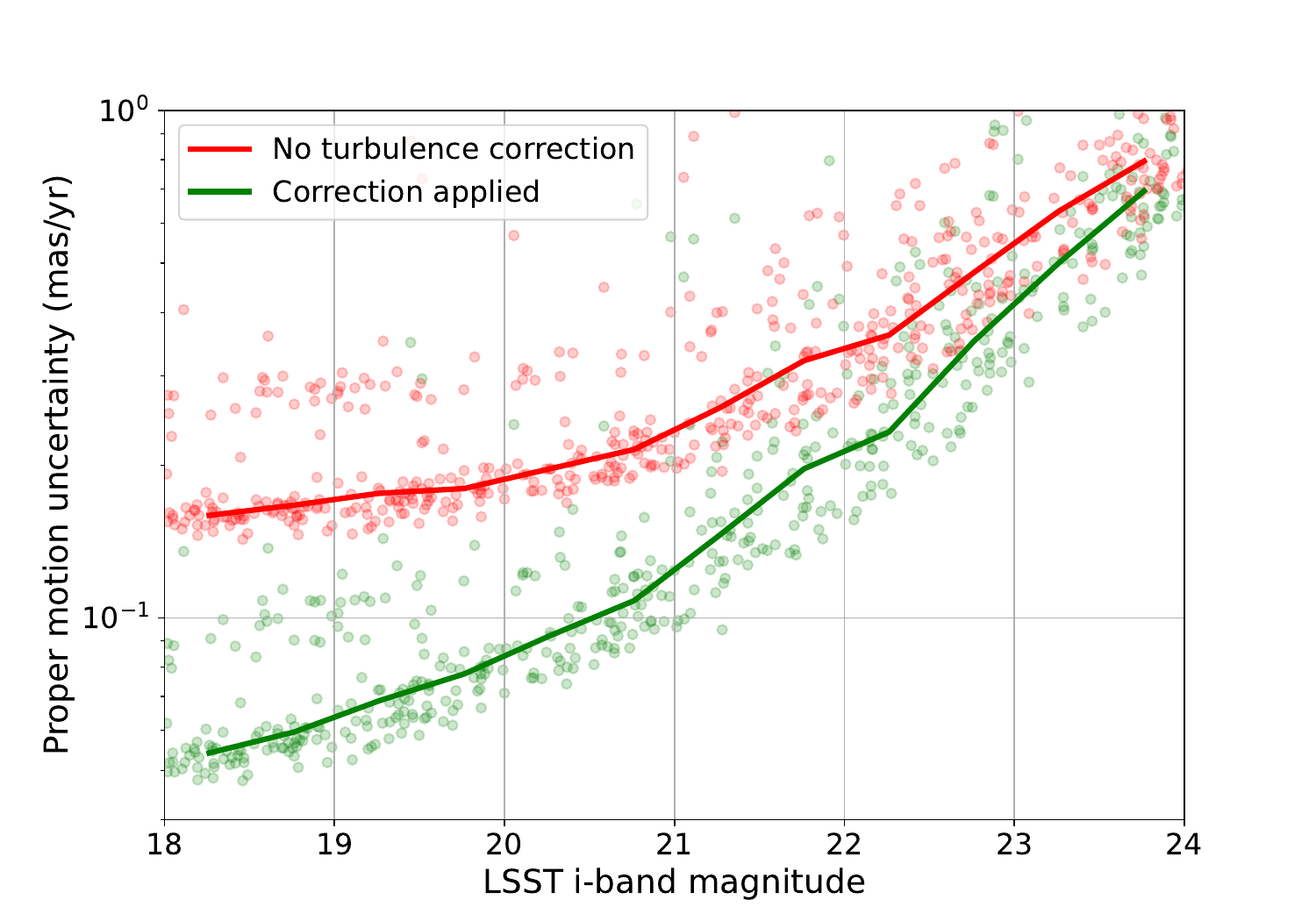}
  \caption{Proper motion errors (in declination)  after 10 years of LSST for a sample of ~1000 randomly drawn sky positions and magnitudes, as a function of the i-band magnitude. Red points represent the scenario where no turbulence correction is applied (10\,mas errors added in quadrature to the photon noise error). Green points reduce the astrometric errors according to the performance of our turbulence reduction code (the astrometric floor is moved from 10\,mas to 3\,mas). Continuous lines are medians.}
  \label{fig:proper}
\end{figure}

Moving to stellar astrometry, we repeat the same uncertainty forecasting procedure to predict positional uncertainties of LSST stars at random magnitudes and sky positions. A photon-noise uncertainty is computed, and combined in quadrature with a 10\,mas floor (uncorrected scenario) or a 3\,mas floor (corrected scenario). We then propagate the uncertainties of all detections of a single star to produce proper motion uncertainty estimates. The proper motion in declination errors for both scenarios are shown in Figure \ref{fig:proper}.

\section{Discussion}
We have produced a code that yields an order-of-magnitude reduction in the variance of astrometric errors from atmospheric turbulence, through Gaussian process interpolation of displacements from \textit{Gaia} DR3 stellar positions. On a set of 604 DES exposures, we obtained an average reduction of $12\times$ in this variance, leaving typical RMS unmodeled displacements of just 2--3\,mas. Our code does not require kernel optimization and makes use of fast correlation measurements, greatly reducing computational cost when compared to \cite{Fortino_2021}. It is now feasible to post-correct DECam exposures in $\sim$\,500 core-seconds, and we expect LSST exposures to require only $2.5 \times$ as much time.

The precision gain achievable with our code would have many further benefits if applied to a survey such as LSST. Improved astrometric precision for solar-system bodies can translate directly into more precise
orbital constraints for a given number of measurements, with secure dynamical understanding
earlier in the survey (identifying interstellar objects as such, and measuring non-gravitational
forces). These orbital constraints also allow small bodies to better serve as test particles for gravitational perturbations such as
the presence of a massive Planet X (as suggested in \citet{Gomes_2023}), and for deflections from mutual asteroid encounters, from which the mass of the deflector can be computed \citep{Bernstein2025}.
Better accuracy in the measurement of stellar parallax and proper motion with Rubin will lead to Gaia-quality catalogs for many stars fainter than the Gaia limit. This will substantially enhance ability to discover and measure velocities for Milky Way dwarfs and streams spanning large angles, and consequent constraints on Galactic substructure.
\begin{acknowledgments}
This work was supported by National Science Foundation grant AST-2205808.  We thank Patricia Burchat for her contribution towards the LSST turbulence simulations and for helpful comments.
\end{acknowledgments}

\bibliography{refs}{}

\end{document}